# Enhancing Exploratory Learning through Exploratory Search with the Emergence of Large Language Models

Yiming Luo
Faculty of Applied Sciences, Macao Polytechnic University
Tacklesss1@gmail.com

Patrick Cheong-Iao Pang
Faculty of Applied Sciences, Macao Polytechnic University
mail@patrickpang.net

Shanton Chang
School of Computing and Information Systems, The University of Melbourne
shanton.chang@unimelb.edu.au

## Abstract

*In the information era, how learners find, evaluate, and effectively use information has become a challenging issue, especially with the added complexity of large language models (LLMs) that have further confused learners in their information retrieval and search activities. This study attempts to unpack this complexity by combining exploratory search strategies with the theories of exploratory learning to form a new theoretical model of exploratory learning from the perspective of students' learning. Our work adapts Kolb's learning model by incorporating high-frequency exploration and feedback loops, aiming to promote deep cognitive and higher-order cognitive skill development in students. Additionally, this paper discusses and suggests how advanced LLMs integrated into information retrieval and information theory can support students in their exploratory searches, contributing theoretically to promoting student-computer interaction and supporting their learning journeys in the new era with LLMs.*

**Keywords:** Exploratory search, Exploratory learning, Large language models, Information retrieval, Learning theory.

## 1. Introduction

With the continuous development of information and communications technology (ICT), especially the increasing richness of digital resources, digital libraries, and Massive Open Online Courses (MOOCs), students at all educational stages now have access to more learning resources than ever before. However, the issue accompanying the information explosion is how students can find, evaluate, and effectively use this information (Dziuban et al., 2013). Exploratory search is a type of information-seeking activity distinct from targeted search, which involves looking for specific goals and expecting specific results (Marchionini, 2006). Exploratory search involves initially undefined and ever-changing information needs (White & Roth, 2009). A classic example of exploratory search strategies is when searchers start with vague or unclear search goals, engaging in exploratory browsing to learn, investigate, and discover relevant information (Pang, et al., 2015). As they become more certain about the topic, they use focused search to locate specific documents and extract the needed information for their tasks. Rieman defines exploratory search as an effective and engaging strategy for learning new systems or investigating unfamiliar features in familiar software. In exploratory search, users do not work through precisely ordered training materials but proactively explore the system, often in pursuit of real or artificial tasks (Rieman, 1996).

One of the reasons that exploratory search is effective is its proven association with other information-seeking behaviors (Savolainen, 2018), such as the Berrypicking and Information Foraging theories, which aim to explain how people search for information. For example, based on the concept of "information scent," information seekers detect and use signals (such as web links or literature citations) to move from one piece of information to another, searching for information relevant to their goals. The Berrypicking model is an information-seeking model where searchers enter the information space and select relevant information dispersed across different documents. Each step in the search brings new ideas to the searcher, potentially redefining queries or search goals and constantly updating information needs. From the perspective of information-seeking behavior, exploratory search requires numerous queries to further understand the topic, explore independent aspects, and respond to emerging information (Golovchinsky et al., 2012).

Although exploratory search has become a popular way to learn, acquire new knowledge, and solve problems (Huang & Yuan, 2024), at present, college students still have various problems and troubles in its application (Bakermans, 2018). In the field of education, researchers tend to believe this is related to information ability and literacy (Irfan et al., 2024). However, it is treating the symptoms: with the addition of search tools such as generative artificial intelligence, students are faced with a constant stream of new information seeking problems and information search skills that need to be constantly upgraded. Therefore, a question worth delving into is how to fundamentally enhance students' ability to conduct exploratory searches to adapt to the ever-changing technological environment?

This paper will address this issue from a new perspective, focusing not on the effectiveness of search behaviors and information retrieval systems but on the concept of student learning. By mapping the concept of exploratory search onto students' exploratory learning theories, this approach will transform exploratory search from an isolated search strategy into an exploratory learning theory that fundamentally improves students' learning methods and thinking. Furthermore, in addition, this paper will also discuss and suggests how students can be supported in exploratory search through state-of-the-art LLMs integrated into information retrieval and information theory.

## 2. Literature Review

### 2.1. Exploratory Learning

Distinct from exploratory search strategies in information retrieval, there is a similar concept in educational theory to represent a more active learning process known as exploratory learning. This emphasizes learners' active exploration, discovery, and reflection to acquire new knowledge and solve problems. It involves iterative learning processes, deep cognitive engagement, and the integration of real-world contexts and social interactions, aiming to cultivate learners' problem-solving skills and critical learning abilities. De Freitas suggests that exploratory learning is a mode of learning structured around associative, cognitive, and situational elements (De Freitas & Neumann, 2009). The Canadian Ministry of Education defines it as the ability to "think critically and creatively through inquiry, reflection, exploration, experimentation, and trial and error, and how to make discoveries" (Alberta Education, 2010). The most classic instructional design theories of exploratory learning include constructivist teaching and Inquiry-Based Learning (IBL). Constructivism posits that learning is a process based on learners' prior knowledge and experiences, involving the interaction and reconstruction of new information to build new knowledge and understanding. Learners construct their understanding and knowledge structure through active participation in problem-solving, discussions, and practical activities (Jonassen, 1991). Inquiry-Based Learning (IBL) is a student-centered learning method that builds a knowledge system through posing questions, conducting investigations, collecting and analyzing data, communicating and sharing, and reflecting and evaluating (Pedaste et al., 2015). Kolb, building on early constructivist and inquiry-based learning, proposed the experiential learning cycle (Kolb, 2014). It operates in a cyclical manner, involving four stages: experiencing, reflecting and observing, forming abstract concepts, and testing. As shown in Figure 1, earners reflect on and observe their experiences after engaging in activities or encountering events, considering the meaning and impact of these experiences, identifying patterns and regularities, and summarizing the reflections. Learners then form abstract concepts and theories, involving the theorization of experiences and the conversion of specific experiences into general principles and concepts, which are applied in new contexts and tested through practice to validate and adjust their understanding and assumptions.

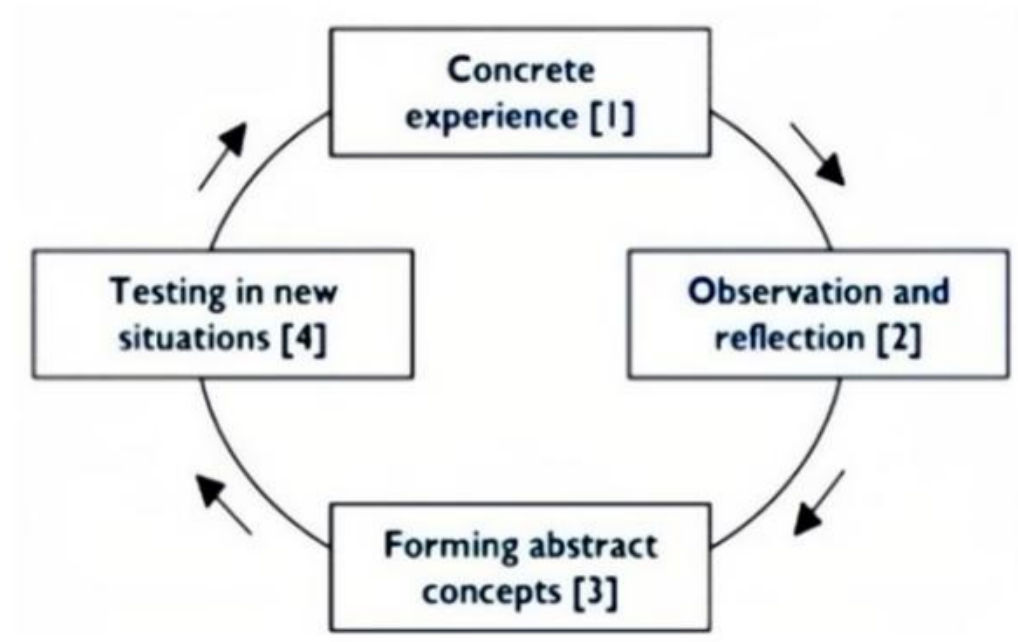

**Figure 1. Kolb experiential learning cycle**

Traditional exploratory learning often faced skepticism, such as Kolb's experiential learning cycle's reliance on external interaction and real-world experiences. In the early stages of educational digitalization, the teaching environment was relatively rigid, information exchange channels were scarce, learning resources were limited, and students' external interactions depended primarily on teachers. Furthermore, constructivist methods have been criticized for neglecting the characteristics of human cognitive structures. A substantial body of empirical evidence suggests that in most cases, instruction emphasizing guidance yields better results (Zhang, 2014). The characteristics of human cognitive structures dictate that effective exploratory teaching requires a certain degree of prior knowledge, which is especially challenging for students who already lack channels for information access. The practical development and implementation of exploratory learning have been influenced by information retrieval.

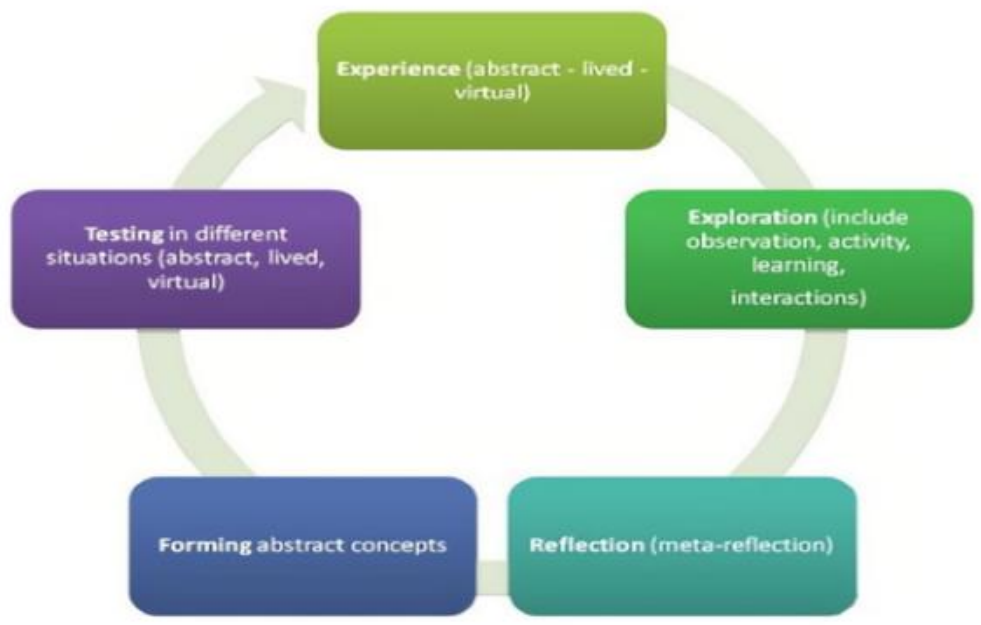

**Figure 2. De Freitas experiential learning cycle**

In the era of educational digitalization, with the development of ICT technology, a series of open standard electronic tools and services are prompting a reconsideration of how, where, and what we learn (Collins & Halverson, 2018). Especially with electronic learning methods such as MOOCs and distance learning, the means for students to obtain external interactions and information have significantly increased. In these enriched virtual environments based on various technologies, learners' roles are empowered, making learners more active and autonomous. De Freitas restructured the original Kolb experiential learning cycle, adding an "exploration" module (De Freitas & Yapp, 2005), as shown in Figure 2. However, a common drawback of this inquiry-based learning model is insufficient recognition of the importance of interaction. Kayes emphasized the importance of dialogue in traditional experiential learning processes (Kayes, 2002). Dialogue was the primary form of interaction in early inquiry-based learning. Additionally, De Freitas' supplementary model often reflects passive behaviors in response to mutations in virtual and physical environments. This indicates that the primary purpose of exploration is to support reflections on the information explosion and technological and environmental changes, yet it lacks support related to learning strategies and information interaction.

## 2.2. The development of large language models in information retrieval

With the explosion of learning resources and information, it has become increasingly difficult for learners to find relevant and valuable information that meets their learning needs in an environment of information overload, leading to inefficiencies in exploratory search. However, the rapid development of deep learning technology in artificial intelligence, especially the emergence of large language models like ChatGPT, has brought new advancements to the application of exploratory search strategies in the learning process. Natural language processing (NLP) is a research direction in artificial intelligence aimed at facilitating interaction between computers and humans through natural language (Jiang et al., 2023). Research by Shaari et al. has shown that, through certain exploratory learning strategies, a philosophical bridge can be drawn between NLP and pedagogy (Shaari & Hamzah, 2016). Early models based on NLP had limited abilities to understand complex language structures, and in early practices, although the synergy between information retrieval and NLP could, in principle, provide a powerful foundation for exploratory search, information retrieval emphasized understanding at a structural and macro level, while NLP focused on understanding at the micro level and structure of language (Manning, 2016). This led to text comprehension issues in exploratory searches using NLP due to semantic understanding challenges.

The transition from statistical language models to neural language models offered a pathway to overcoming these challenges. In 2003, Y. Bengio proposed the neural network language model based on deep learning technology (Bengio et al., 2003). This kind of text-related issue can cause learners to encounter more erroneous information and confusion. However, deep learning technology, through training multi-layer network structures, can capture deeper abstract semantics of text and images. The model can better understand users' search intents and extract relevant and valuable content from massive amounts of information. This helps address the problem of information overload, thereby enhancing the efficiency and effectiveness of exploratory search. For instance, using NLP technology enables the establishment of new educational platforms that automatically compile information from various sources, which greatly reduces the time spent by teachers and students in searching for relevant and useful information on various topics (Montalvo et al., 2018).

The emergence of LLMs has brought a further revolution in semantic understanding. The publication of "Attention is All You Need" in 2017 marked a significant turning point in the field of language modeling (Vaswani et al., 2017). The Transformer architecture, which includes a self-attention mechanism, enables the model to weigh the importance of each word in a sequence relative to every other word. Neural language models based on the pretrained Transformer architecture on a massive corpus demonstrated exceptional capabilities in a variety of natural language processing tasks. A key insight from this transformation was the profound impact of scaling model architectures and data on overall performance and functionality. As researchers scaled the parameters and training regimes of these models, they observed improvements in model capabilities, as well as the emergence of specialized abilities, such as contextual understanding. The emergence of LLMs resulted from this shift.

LLMs are state-of-the-art artificial intelligence systems designed to process and generate text, with a focus on coherent communication. The distinction between large models and typical artificial intelligence models lies in LLMs' massive parameter scale. Trained in a self-supervised environment on extensive text corpora, they learn complex language patterns and structures, enabling a qualitative leap in performance in translation, summarization, information retrieval, and natural dialogue interactions, among others (Liu et al., 2021). Furthermore, LLMs possess the capability to integrate information from various sources, providing comprehensive recommendations (Min et al., 2023). At the same time, driven by prompting strategies such as chain-of-thought, the LLMs can generate outputs with step-by-step reasoning to guide complex decision-making processes (Wei et al., 2022). Leveraging the powerful capabilities of LLMs can undoubtedly extend the depth and breadth of exploratory search. By integrating these language models with information systems, the paradigms of information retrieval and exploratory search will be reshaped.

### 2.3. Students' struggles in information search

Numerous studies indicate that students seem to struggle with this capability of using search to learn and solve problems. For instance, students often exhibit blind confidence in search results (Douglas et al., 2014), cognitive biases in evaluating information (Besharat-Mann, 2024), and difficulties in appropriately applying these strategies in specific search scenarios (Hoeber & Storie, 2022). Students tend to have a mechanical understanding of exploratory search methods without deeply comprehending the logic behind them. Furthermore, students lack a thorough understanding of the relationship between search and learning. Although the concept of "search as a learning process to support and improve human learning" was proposed long ago (Rieh et al., 2016), However, there is still a lack of research linking the two fields. Coincidentally, although researchers have developed numerous exploratory search strategies, evaluating exploratory search systems is still considered a difficult and nuanced activity. It requires both qualitative and quantitative analysis of both user behavior and search output. These are complicated systems that combine various features and behaviors, creating an alchemy that is not straightforward to evaluate (White et al., 2008). For example, some exploratory search systems use standard precision and recall IR metrics or more specific metrics such as query length, maximum scroll depth, and task completion time (Athukorala et al., 2016), while some studies may use the questionnaire-based System Usability Scale for evaluation (Liu et al., 2022). It is reasonable to speculate that these inconsistent evaluation standards may lead to students feeling lost and unsure when trying to internalize exploratory search strategies as learning strategies.

As generative AI is becoming more and more popular and widely used by students in their daily learning, the above problems become more serious and urgent. Although exploratory search systems based on LLMs have contributed greatly to the rate of information retrieval, problems such as fabrication of seemingly plausible but incorrect or anachronistic information (often referred to as "hallucinations") pose a significant mindset and moral risk to students (Zhuo al., 2024). In addition, students' over-reliance on this tool has hurt their ability to think critically, explore and summarize on their initiative, greatly affecting their learning outcomes and development (Kasneci et al., 2023). Therefore, there is an urgent need to systematically sort out the relationship between exploratory search strategies and the learning process from the ground up, to properly guide students to rationally utilize a range of enhanced exploratory search tools, including LLMs, to increase students' critical thinking and independent learning skills, and problem-solving abilities.

## 3. Methodology

To investigate the relationship between exploratory learning and exploratory search strategies, we performed a literature analysis with Citespace and LDAvis to visualize the literature keywords and abstracts in the past ten years and aimed to explore the connection between these two concepts to find new perspectives and promote interdisciplinary research. CiteSpace (Chen, C & Chen, C, 2003) is a visualization tool for analyzing scientific literature, mainly used to explore developments and cutting-edge trends in different fields. LDAvia (Sievert, C., & Shirley, K. 2014) is a text-based topic analysis tool that automatically identifies and extracts latent representations in text.

Specifically, relevant literature was retrieved from the Web of Science database. Using "Exploratory Search" and "Exploratory Learning" as keywords, the top 200 relevant articles from 2014 to 2024 were searched, and after removing duplicate articles, a total of 184 relevant papers were retrieved. These documents were imported into CiteSpace to explore the research trends and interrelationships in these two fields. The specific settings were as follows: time slices from 2004 to 2024, one slice per year; node type as keywords; selection criteria as the Top 50 high-frequency keywords; visualization settings using Cosine similarity measurement, and the link retention factor as 2.5 times the number of nodes. To find deeper internal links in the field of education, especially in the sub-field of learning theory, the WOS search strategy is readjusted, initially using "Exploratory Search" and "education" as keywords to collect the top 100 most relevant papers, and then using "Inquiry-Based Learning" and "education," more refined and specific educational field keywords, to repeat the same collection operation. After that, LDAVis was used to perform topic modeling on the abstracts of the articles to measure the connections between the subfields within the field of education in a more quantitative way.

## 4. Analysis and finding

The results were analyzed to identify past connections and potential cross-disciplinary conceptual interactions between exploratory search and exploratory analysis. As shown in Figure 3(a), the results of the keyword clustering analysis indicated that "Exploratory Search" and "Exploratory Learning" formed independent clusters (#0 and #3), suggesting that these two concepts have few connections in the existing literature. Additionally, other main clusters, such as #1 Knowledge Base, #2 Exploratory Product Search, and #4 Exploratory Motivation, also showed their independence. The link found from the keyword clustering analysis is the #1 Knowledge Base cluster. The papers in this cluster mainly cover knowledge acquisition, storage and management research, especially in information systems and knowledge discovery. This may indicate that both the concepts of exploratory learning and exploratory search involve knowledge acquisition, information processing, and iterative feedback. The existence of this cluster seems to suggest that further attention should be given to the knowledge cycle and iteration in exploratory search and exploratory learning. Burst word detection revealed 16 keywords with the strongest bursts, as shown in Figure 3(b). These keywords displayed burst characteristics within specific periods. For example, "Exploratory Learning" showed a strong burst during 2007-2016, while "Exploratory Search" displayed a burst during 2017-2018. Although both exhibited bursts in certain years, their burst periods mostly did not overlap, further supporting the independence of the two concepts in the literature.

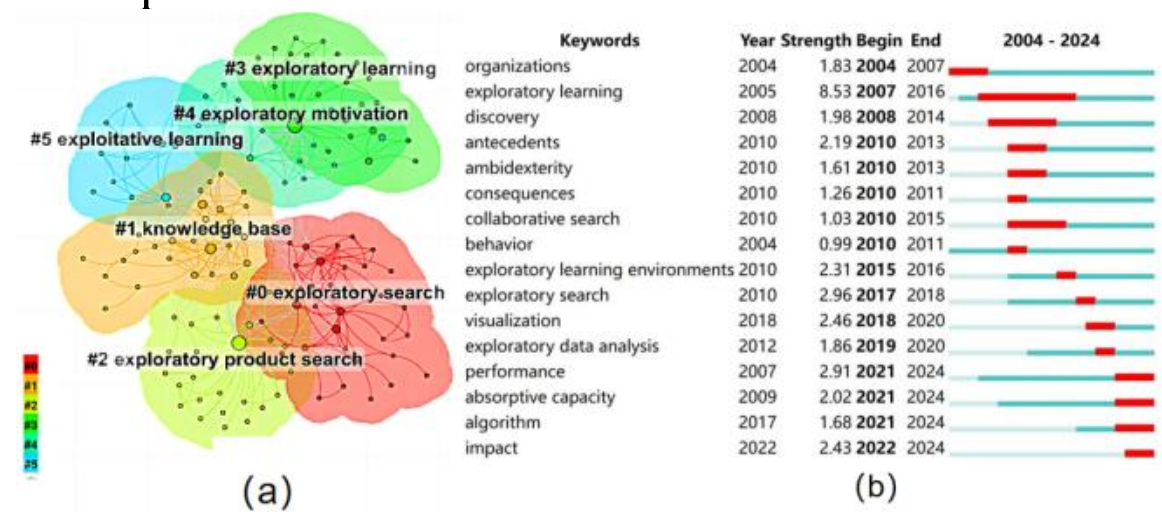

**Figure 3. (a) Keyword clustering and (b) keyword mutation detection)**

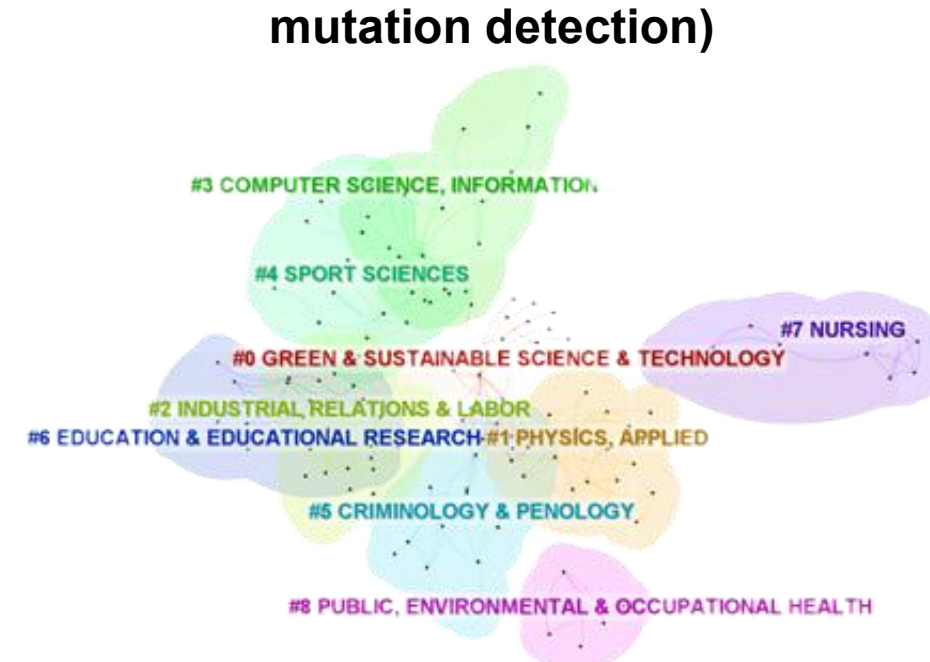

**Figure 4. Subject clustering**

It is further found that both the concepts of exploratory search and exploratory learning pointed mostly to the fields of information systems and computing in the clustering, while content related to the field of education was minimal. The keywords in the obtained nodes also support this point. As shown in Figure 4, the results of the subject clustering of the literature's keywords clearly show that there is no overlap between research in the education field and research in the computing and information retrieval fields. Additionally, using LDAvis as a tool to visualize topic modeling on the abstracts of all the articles again revealed that when re-analyzing the abstracts' topic modeling, none of the topics overlapped, and the topic emphasizing students' exploratory information retrieval and search behavior was the farthest from other topics in Figure 5. All these pieces of evidence suggest that research related to learning theory in the field of education has not been closely linked with knowledge in the fields of computing and information retrieval. Therefore, exploratory retrieval strategies may not be deeply applied to students' learning process, resulting in theoretical gaps and deficiencies in students' information exploration learning process.

The question of how the characteristics of exploratory search strategies is integrated into the learning process requires an understanding of the changing learning environment. With the development of ICT technology, traditional places for information interaction have shifted from offline, fixed classroom models to hybrid, open online learning platforms and information retrieval systems. Information retrieval systems provide users with initial, broad information sources, upon which exploratory search conducts deeper exploration and understanding (White, 2016). In information retrieval, exploratory search typically requires multiple iterative searches. After obtaining initial information, learners adjust their search queries based on new discoveries, conducting further searches. It requires cognitive processing and interpretation to continuously optimize search strategies and results (Zhang, 2014). It involves in-depth analysis and evaluation of information, requiring complex and specialized information integration skills to achieve multidimensional, multimodal human-computer interaction (Ding et al., 2024).

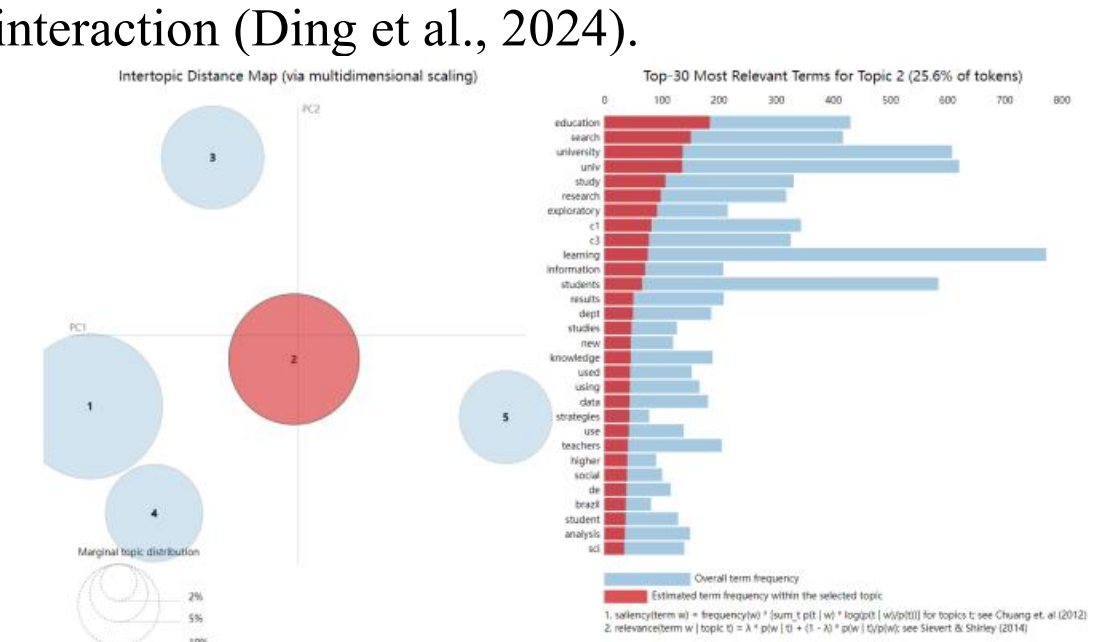

**Figure 5. Abstract topic analysis using LDA visualization**

We found that exploratory learning enhanced by exploratory search strategies has two characteristics that are not present in traditional exploratory learning processes. This disparity is also indicative of the disconnect between traditional learning theories and the modern technological era. As shown earlier in Figure 5, one of the core characteristics of exploratory search is the high frequency of internal and external interactions. The traditional exploratory learning process is usually goal-oriented or divergent, involving fewer non-linear iterations and adjustments (Kirschner et al., 2006). However, in the learning process guided by the exploratory search strategy, learners need to conduct a series of complex, non-goal-oriented searches between information retrieval systems and provide feedback, re-entering the internal exploration loop. Through these iterative exploration and feedback cycles, learners can conduct deeper analysis and evaluation of information. Another core characteristic is the deep exploration and feedback loop. Although traditional inquiry-based learning does not usually rely on highly structured evaluation models to measure school effectiveness, it still focuses on phased conceptual results (Ramnarain, 2023). Compared to the traditional exploratory learning process on the left side of Figure 6, exploratory search allows learners to move to the next exploration step after each feedback without having to form new concepts immediately. It provides learners with more space to explore deep information and helps them to continuously adjust and optimize their cognitive framework during the exploration process.

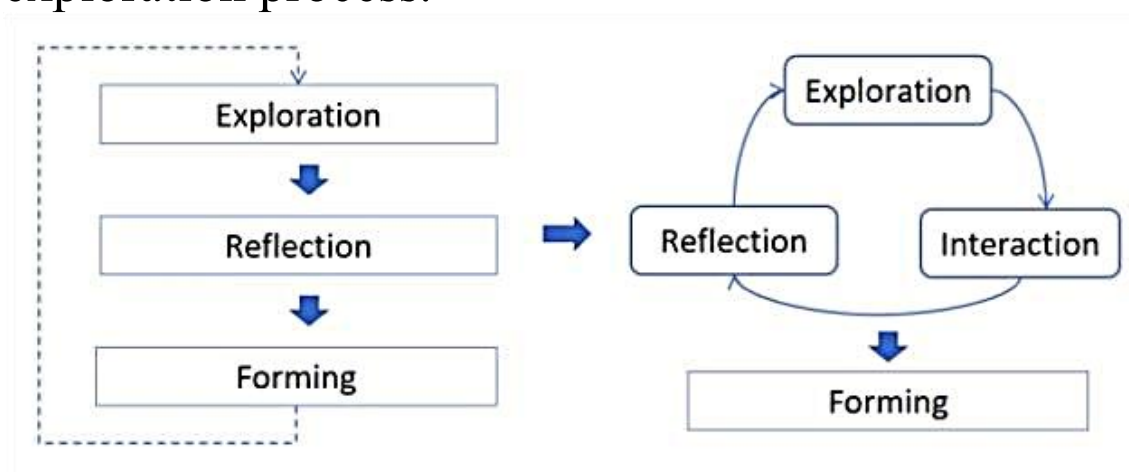

**Figure 6. Improved exploratory learning process based on exploratory search strategy**

Exploratory search plays a crucial role in developing higher-order cognitive skills. According to White (White, 2016), higher-order cognitive skills include creation, evaluation, and analysis capabilities. Based on Bloom's taxonomy, as discussed by Muhayimana et al. (Muhayimana et al., 2022), Exploratory search enables learners to continuously access and process new information through a high-frequency exploration and reflection process, helping them to connect new information with existing knowledge and form a deeper understanding. Through continuous reflection, learners can internalize knowledge and transform it into understanding and practical application, which is very important for the development of higher-order cognitive levels. (Marton & Säljö, 1976). Some practices further prove that high-frequency interactions can significantly enhance students' conceptual understanding and task execution capabilities, thereby providing strong support for developing higher-order cognitive skills (Lee, 2023).

It should be noted that this high-frequency cycle only exists in human-computer interaction and does not involve the process of students' self-reflection. This means that to cultivate students, more exploration and iteration are needed before hastily forming concepts, rather than quickly engaging in self-reflection cycles. In the information age, when knowledge in the external environment changes rapidly, quick self-reflection cycles may lead to delayed decision-making and the inability to take timely actions. This is reflected in students' confusion when faced with search tasks. On the other hand, quick self-reflection cycles may continuously reinforce existing concepts and methods, thereby limiting the acceptance of new information and methods. This reflects students' blind confidence in search results and methods. Overall, in the new era of information technology development, to form a good exploratory learning process, it is necessary to borrow high-frequency exploration feedback cycles from exploratory search strategies. Therefore, increasing the frequency of human-computer interaction, accelerating the exploration iteration process, and appropriately relaxing the pace of reflection can help students avoid cognitive biases, overcome the problem of being stuck in local optimal solutions, and truly improve their problem-solving abilities.

## 5. Discussion

### 5.1. How to train students to be more exploratory in the era of LLMs

As LLMs emerged, the paradigm of information acquisition for learners gradually shifted from using IR systems to search for information to using LLMs to generate information. Besides, exploratory search requires special design to support contexts (Pang et al., 2016). In this paradigm shift, IR transitioned from directly assisting humans to indirectly assisting LLMs. This is because, unfortunately, traditional information retrieval systems, due to their architectural limitations, cannot fully utilize the powerful semantic understanding capabilities provided by LLMs. Despite efforts to enhance IR systems by integrating LLMs, these methods typically only leverage shallow representations of LLMs to fine-tune information retrieval models (Ma et al., 2023) or make limited improvements to specific parts of IR systems based on LLMs (Pradeep et al., 2023). However, the appearance of end-to-end large model search systems (Tang et al., 2024) and large model agent systems (Mei et al., 2024) in recent years indicates that information retrieval based entirely on generative large models is gradually becoming the new mainstream. This has also led to subtle changes in the application of exploratory search in the learning process. Although the core influence of exploratory search on the learning process, namely multiple iterative queries, exploration, and feedback loops, has not changed, the human-computer interaction relationship has undergone a dramatic change. Learners have transitioned from interacting with traditional information retrieval platforms and non-artificial intelligence computing systems to interacting with information retrieval systems based on LLMs. They resemble high-dimensional intelligence rather than tools. They can achieve context-aware interactions to understand the nuances of human language and the context in which interactions occur, providing personalized content recommendations and value output to learners. As shown in Figure 7(a), unlike traditional methods that clarify user search requests and preferences and provide explicit relevance feedback. LLMs will interact with learners in a more implicit and equal relationship. In the process of exploratory learning, LLMs will act on the learners' exploration and feedback processes in a more implicit and equal relationship, forming an invisible interactive cycle. This means that LLM-based information retrieval systems will no longer be rigid external devices but intelligent entities with general thinking capabilities (Leng & Yuan, 2023).

Correspondingly, the role of learners should also change in the era of LLMs. Students are no longer merely inquirers, but more like managers and agents. Students are responsible for setting research goals and subdividing directions, while LLMs are responsible for executing and providing information. Thus, the human-computer interaction between learners and LLMs is more like the collaboration between managers and grassroots workers. Therefore, when evaluating learners' problem-solving abilities, more attention should be paid to their exploratory learning process using LLMs, including their thinking methods, problem decomposition abilities, and adjustment abilities during the exploration process, rather than merely the structured or unstructured results. For example, when faced with a task such as learning a new language, a vague study plan or a detailed timetable may no longer be the desired answer in the era of large models. Instead, finding and evaluating the experiences and theories used by many successful language learners, assessing one's own learning characteristics, matching and evaluating these with successful experiences, and ultimately forming an execution process is a blueprint that can reflect the afore-mentioned abilities. This is precisely what can be achieved through an exploratory learning process based on exploratory search. In other words, LLMs can serve as a catalyst in the implementation of exploratory search strategies and the exploratory learning process. LLMs significantly simplify the process by which students conduct exploratory searches, providing a highly effective and promising interactive approach that enables students to engage in exploratory learning more easily and efficiently.

## 5.2. Barriers to effective exploratory learning using LLMs

Applying LLMs for exploratory learning remains controversial. More implicit and closer human-computer interaction does not necessarily have any drawbacks to the learning process. Due to their broad pre-training database and improper parameter tuning, there is a tendency to introduce a lot of irrelevant information due to much detailed and redundant content when applying LLMs, known as concept drift. While this can truly enrich queries to enhance exploratory search, there is also the risk of generating irrelevant or deviating results (Perconti & Plebe, 2020). Furthermore, LLMs may generate inaccurate "hallucinate" information when performing tasks that require obscure real-world knowledge, making it necessary to evaluate the factual accuracy of the model-generated statements. Traditional search systems optimize for exploratory search strategies and typically integrate various techniques such as document clustering and intelligent content summarization, resulting in users encountering numerous irrelevant but potentially enlightening results. Learners must sift through these results to find accurate and relevant information, a process that can be very time-consuming but is still predictable and controllable. LLMs have reimagined traditional search systems. Aside from the initial retrieval stage, all IR tasks are formulated as text generation problems and handled by a single large search model. The large search model automatically generates various elements that constitute the search results page, including ranked document lists, snippets, direct answers, and more (Wang et al., 2024). While the exploratory learning process becomes easier, it will make it more difficult to filter information and trace the source if hallucinations occur and takes extra effort to further correct and explore. Although some techniques such as Retrieval-augmented Generation (RAG) that use external knowledge bases to supplement the context of LLMs and generate responses have been able to alleviate the hallucination problem to a certain extent, human review should be still further strengthened to ensure the reliability and accuracy of the final generated content. Many researchers have proposed various strategies to strengthen methods: Lujain et al. proposed that a new round of human-computer interaction evaluation is needed to re-examine the process and results of human-computer interaction or human use of models. Ji et al. proposed an interactive self-reflection method that combines knowledge acquisition and answer generation (Ji et al., 2023). We benefited from them and tried to incorporate the idea of strengthening reviews into our exploratory framework. As shown in Figure 7(b), we revise the exploratory learning process specifically for LLMs: learners need to evaluate and check the effectiveness of human-computer interaction after exploratory searches. Before forming new concepts and applications, it is essential to introduce external feedback from authoritative figures such as educators or domain experts. This further ensures the elimination of negative impacts from deep interactions with LLMs, thereby guaranteeing that students' exploratory learning remains on the right way. Moreover, the evaluation standards of exploratory search systems remain quite chaotic and unclear. Researchers have adopted methods such as tracing the source (Irving et al., 2018) and establishing adversarial networks for AI debate (Chakraborty et al., 2017) to improve factual accuracy and alleviate the trust crisis. Despite these efforts, progress in achieving interpretable LLM-driven exploratory search systems remains very slow. As a deep and complex black-box model, LLM poses significant challenges to existing model interpretability techniques. This still requires artificial intelligence and education experts to further explore exploratory search systems and exploratory learning processes based on LLM technology and the empirical research sections currently missing due to space limitations.

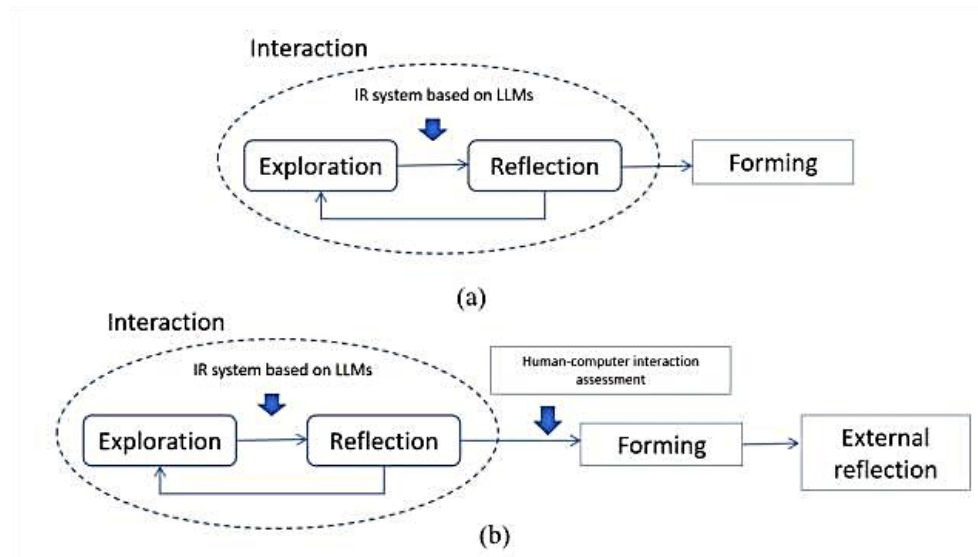

**Figure 7. Comparison of (a) exploratory learning process with LLMs; and (b) with external intervention addressing potential challenges of LLMs**

## 6. Conclusion

With the advent of the information age, effectively finding, evaluating, and utilizing the overwhelming amount of information for learning and reflection have become challenging issues for learners. Despite the immense potential of exploratory search strategies in education, there is a scarcity of research that combines the advanced concepts of exploratory search with learning theories in the educational field. e. This study attempts to address this gap by integrating exploratory search strategies from information retrieval into the learning process from the student learning perspective. This approach aims to form a novel exploratory learning model that combines exploratory search with learning theories. The model adapts Kolb's learning model to the information age, focusing on high-frequency exploration and feedback loops, which help students explore new information under uncertainty and promote deep cognitive and higher-order cognitive skill development. This paper also discusses the opportunities and challenges of exploratory learning theory in the era of LLMs. In future research, we will further strengthen the relationship between exploratory search and the learning process through more examples to verify the accuracy of the exploratory learning model we proposed. In addition, through interdisciplinary research and practice, we can promote the common development of educational theory and information retrieval theory, and ultimately achieve the goal of improving students' problem-solving ability in the information age.

## 7. Acknowledgements

This research was funded by Macao Science and Technology Development Fund (funding ID: 0088/2023/ITP2) and Macao Polytechnic University research grant (project code: RP/FCA-10/2022).